\documentclass[onecolumn,manuscript,showpacs,amsmath,amssymb]{revtex4}
\usepackage{graphicx}%
\usepackage{dcolumn}
\usepackage{amsmath}
\usepackage[dvips]{epsfig}

\makeatletter
\def\btt#1{\texttt{\@backslashchar#1}}%
\DeclareRobustCommand\bblash{\btt{\@backslashchar}}%
\makeatother

\begin{document}

\title{Superconducting Properties of Two-Orbital t-t$^{'}$-J-J$^{'}$
       Models}
\author{Feng Lu$^{1,2}$ and Liang-Jian Zou$^{1,
\footnote{Correspondence author, Electronic mail: zou@theory.issp.ac.cn}}$}
\affiliation{\it
 $^1$ Key Laboratory of Materials Physics, Institute of Solid State Physics,
      Chinese Academy of Sciences, P. O. Box 1129, Hefei 230031, China}
\affiliation{\it
 $^2$ Graduate School of the Chinese Academy of Sciences, Beijing 100049,
      China}
\date{09/02/2008}
\begin{abstract}
   Motivated by the recent contradiction of the superconducting pairing
symmetry in the angle-resolved photoemission spectra (ARPES) and the
nuclear magnetic resonance (NMR) data in the FeAs superconductors,
we present the theoretical results on the phase diagram, the
temperature dependent Fermi surfaces in normal state, the ARPES
character of quasiparticles and the spin-lattice relaxation
1/T$_{1}$ of the two-orbital t-t$^{'}$-J-J$^{'}$ models. Our results
show that most of the properties observed in iron-based
superconductors could be comprehensively understood in the present
scenario qualitatively, indicating that the pairing symmetry of the
ironpnictides is anisotropic nodeless $d_{x^{2}-\eta
y^{2}}$+$S_{x^{2}y^{2}}$-wave,  mainly originating from the band
structures and the Fermi surface topology.

\end{abstract}


\maketitle

\section{INTRODUCTION}

Following the discovery of the first high temperature
superconductivity in the copper-based compounds two decades ago
\cite{Bednorz}, a second class of high temperature superconductors
has been recently reported in iron-based pnictides
\cite{Kamihara08,Ren,GFChen1,XHChen,GFChen2,HHWen}, in which the
transition temperature $T_{c}$ can be as high as $55 K$ \cite{Ren2}.
Intensively experimental \cite{Chen,la_Cruz_NeutronScattering_SDW,
Hunte,Jia_PE,Liu_PE, FENG,CHANGLIU1,CHANGLIU2,
Hashimoto_FullGap,YangBaK_ARPES,LiuSrK_ARPES} and theoretical
\cite{Mazin,Kuroki,Xi,PALee,Yao,LeeDH} efforts have been devoted to
understand the nature of new superconductors.
Resemble to cuprates, layered iron pnictides consist of the
conducting FeAs layers and the ReO layers (Re represents the
rare earth elements, such as La, Ce, Pr, Nd, Sm, and $etc$) which
provide carriers to the FeAs layers. The antiferromagnetic spin
density wave (SDW) order is suppressed upon substituting a few
percent O with F, and the new compounds become the superconductor
(SC) below T$_{c}$.
On the other hand, there are a few of considerable differences
between cuprates and ironpnictides. For example, the first
difference is that the undoped iron-oxypnictides with the SDW order
exhibit poor metallic conduction, rather than the {\it Neel} AFM
insulator in undoped cuprates; the second one is that the spin and
magnetic moment of Fe are much smaller than expected, completely
different from the Cu spin in cuprates, {\it etc}. These facts
suggest that there exists significant difference between the ground
states of ironpnictides and cuprates.

First-principle electronic structure calculations provide the first
evidence of the difference. It has been shown
\cite{10,11,12,Cao,14,15}
that in LaFeAsO, the most of the spectral weight close to the Fermi
energy is contributed from the Fe $3d$ orbital, and the Fermi
surfaces (FS) of LaFeAsO consists of two hole-type circles around
$\Gamma$ point and two electron-type co-centered ellipses around M
point \cite{ZhangHJ}. This implies that the multiorbital character
is dominant in the undoped FeAs superconductor, contrast to the
single-orbital character in cuprates.
Up to date, many tight-binding multi-orbital models have been
proposed to reproduce the Fermi surface (FS) character and the band
structures near Fermi level E$_{F}$ in ReFeAsO$_{1-x}$F$_{x}$ and
Ba$_{1-x}$K$_{x}$Fe$_{2}$As$_{2}$.
To this end, once the electron-phonon mechanism is precluded by
Boeri {\it et al.} \cite{Boeri}, the electronic mechanisms are
proposed as the driving force of the SC paring, such as the Coulomb
interaction between Fe $3d$ electrons \cite{PALee,Nomura,Xi,LeeDH}
or the antiferromagnetic exchange coupling between the nearest
neighbor Fe sites and the next nearest neighbor site, such as the
two-orbital models \cite{Han,Raghu,Baskaran}, three-orbital ones
\cite{PALee}, and four-orbital ones \cite{Korshunov}, even the
five-orbital model \cite{Kuroki,Haule1}.
Among these models, the two-orbital t-t$^{'}$-J-J$_{'}$ model
\cite{Han,Raghu} is minimal, and can reproduce the key characters of
ironpnictides, such as the complex Fermi surface and the
multi-orbital degeneracy of the Fe $3d$ electrons \cite{Han,Raghu}
and the band structures near E$_{F}$, as well as the stripe
antiferromagnetic or spin-density-wave ground state.

In this newly discovered superconductor, the most intriguing issues
are the pairing mechanism and pairing symmetry. To date, various
possibilities of the SC mechanism and pairing symmetry for
$Fe-$pnictide superconductor are proposed theoretically
\cite{Seo,PALee,Nomura,Han,Xi,LeeDH} and experiemntally
\cite{Kondo,Zhou,Ding,Matano,Ahilan}.
The pairing symmetries range from spin singlet $d$-wave
\cite{Wen2,Millo} to $s$-wave\cite{LeeDH,Nomura}, or the mixture of
$S_{x^{2}y^{2}}$ and $d_{x^{2}-y^{2}}$ \cite{Seo}, or spin triplet
$p$-wave\cite{Xi,PALee}.
At present, the nature of the SC gap observed experimentally is very
different from authors to authors and from samples to samples,
ranging from one gap \cite{Chen} to two gaps
\cite{Wang2,Matano,Ding}, and from isotropic \cite{Ding,Zhou} or
anisotropic fully gap \cite{Kondo} to line node gap \cite{Matano}.
On the one hand, the angle-resolved photoemission spectroscopy
(ARPES) measurements \cite{Ding,Zhou,Kondo}, a direct measurement to
the quasiparticle spectra and the SC gap, observed that all the gaps
are nodeless around their respective Fermi surface sheets, which is
different from the situation of the cuprates.
On the other hand, however, the spin-lattice relaxation rate in the
nuclear magnetic resonance (NMR) experiments strongly suggest the
existence of nodes in the gap \cite{Matano}, which share partial
features of cuprates and MgB$_{2}$.
The question thus arises whether the contradiction between the ARPES
and the NMR data is due to the sample qualities or it reflects the
two sides of the new superconductors ?

In this paper, we start from the minimal two-orbital
t-t$^{'}$-J-J$^{'}$ model \cite{Seo}, which has the same topology as
the band structure of the iron based superconductors, and obtain the
mean-field phase diagram of the extended t-t$^{'}$-J-J$^{'}$ model,
the quasiparticle spectra and its ARPES manifestation, and the
spin-lattice relaxation rate 1/T$_{1}$. Our results demonstrate that
the various pairing symmetries are stable in the mean-field phase
diagram, and the T$^{3}$-like behavior in the spin-lattice
relaxation rate may coexist with the anisotropic nodeless SC gap, as
observed in ARPES and NMR experiments.

\section{Model Hamiltonian and Method}

Based on the band structures results and theoretical analysis, the
twofold-degenerate d$_{xz}$/d$_{yz}$ orbits are essential for the
ironpnictide superconductors. We depict such physical processes with
the two-orbital Hubbard model,
\begin{eqnarray}
  \hat{H}&=& \sum_{<ij>\alpha\beta\sigma} t_{ij}^{\alpha\beta}
     \hat{c}_{i\alpha\sigma}^{\dagger}\hat{c}_{j\beta\sigma}+
           \sum_{\ll ij\gg \alpha\beta\sigma} t_{ij}^{'\alpha\beta}
     \hat{c}_{i\alpha\sigma}^{\dagger}\hat{c}_{j\beta\sigma}
        \nonumber\\
  &&+U\sum_{i{\alpha}}
  \hat{c}_{i\alpha\uparrow}^{\dagger}\hat{c}_{i\alpha\uparrow}
 \hat{c}_{i\alpha\downarrow}^{\dagger}\hat{c}_{i\alpha\downarrow}
 +U'\sum_{i\sigma\sigma'}\hat{c}_{i1\sigma}^{\dagger}\hat{c}_{i1\sigma}
 \hat{c}_{i2\sigma'}^{\dagger}\hat{c}_{i2\sigma'}
            \nonumber\\
  &&-J_{H}\sum_{i\sigma\sigma'}\hat{c}_{i1\sigma}^{\dagger}\hat{c}_{i1\sigma'}
  \hat{c}_{i2\sigma'}^{\dagger}\hat{c}_{i2\sigma}
  +J_{H}\sum_{i\alpha \neq \alpha'}
  \hat{c}_{i\alpha\uparrow}^{\dagger}\hat{c}_{i\alpha'\uparrow}
  \hat{c}_{i\alpha\downarrow}^{\dagger}\hat{c}_{i\alpha'\downarrow}
\end{eqnarray}
here $c^{\dagger}_{i\alpha\sigma}$ creates a $d_{xz}$ ($\alpha$=1)
or $d_{yz}$ ($\alpha$=2) electron with orbital $\alpha$ and spin
$\sigma$ at site R$_{i}$. t and t$^{'}$ denotes the hopping
integrals of the nearest-neighbor (NN) and the next-nearest-neighbor
(NNN) sites, respectively. U, U$^{\prime}$ and J$_{H}$ are the intra-orbital,
inter-orbital Coulomb interactions and the Hund's coupling.
Formally, the t-t$^{'}$-J-J$^{'}$ model can be derived from Eq.(1)
in the atomic limit \cite{Manousakis}, although far from strict.
%
%
The t-t$^{'}$-J-J$^{'}$ model reads,
\begin{equation}
    H=H_{t-t^{'}}+H_{J-J^{'}}
\end{equation}
with the kinetic energy term,
\begin{equation}
   H_{t-t^{'}} = \sum_{k {\sigma}}[ (\varepsilon_{k xz} - \mu)
 c_{k 1 \sigma}^{\dagger}c_{k 1 \sigma}
 +(\varepsilon_{k yz}-\mu) c_{k 2 \sigma}^{\dagger}c_{k 2 \sigma}
  \nonumber \\
  +\varepsilon_{k xy}(c_{k 1 \sigma}^{\dagger}c_{k 2 \sigma}
 +c_{k 2 \sigma}^{\dagger}c_{k 1 \sigma})]
\end{equation}
with the notations
\begin{eqnarray}
   \varepsilon_{k xz} &=& -2t_1\cos k_x-2t_2\cos k_y-4t_3\cos k_x\cos k_y,
  \nonumber\\
   \varepsilon_{k yz} &=& -2t_2\cos k_x-2t_1\cos k_y-4t_3\cos k_x\cos k_y,
  \nonumber\\
   \varepsilon_{k xy} &=& -4t_4\sin k_x\sin k_y.
   \nonumber
\end{eqnarray}
Here the components of t$^{\alpha\beta}_{ij}$ are
t$^{11}_{x}$=t$_1$=-1 and t$^{22}_{x}$=t$_2$ = 1.3, and these of
t$^{\prime \alpha\beta}_{ij}$ are $t_3=t_4=-0.85$ \cite{Seo}.
%
%
The electron is filled with n=2 (or half-filling) on a
two-dimensional square lattice.
The interaction term reads,
\begin{eqnarray}
  H_{J-J^{'}} &=& J \sum_{<ij> \alpha\beta}
    (\vec{S}_{i \alpha} \cdot \vec{S}_{j \beta}
       -\frac{1}{4} n_{i \alpha}\cdot n_{j \beta})
      + J^{'}\sum_{<<ij>> \alpha\beta}
       (\vec{S}_{i \alpha} \cdot \vec{S}_{j \beta}
     -\frac{1}{4}n_{i \alpha}\cdot n_{j \beta})
\end{eqnarray}
where $J$ and $J^{'}$ are the NN and the NNN antiferromagnetic
couplings, through the concrete values of $J$ and $J^{'}$ are still in
debate. Haule $et ~al.$ estimated that $J/J^{'}\thicksim 2$
\cite{Haule2}. In this paper, we assume $J$=0.7 and $J^{'}$=0.3. $\vec
S^{\alpha}_i$ is the spin operator of the electron in the
$\alpha$-orbit, and $n_{i \alpha}$ is the particle number operator.
The orbital indices $\alpha$ and $\beta$ run over 1 and 2.

Introducing the following order parameters,
\begin{eqnarray}
  \Delta^{1 m}_{x/y}&=&
  J<c^{\dagger}_{i m \uparrow} c^{\dagger}_{j m \downarrow}>,
         ~~ (j=i \pm \hat{x}/\hat{y})
    \nonumber\\
  \Delta^{2 m}_{x \pm y}&=&
  J^{'}<c^{\dagger}_{i m \uparrow} c^{\dagger}_{j m \downarrow}>,
        ~~(j=i \pm \hat{(x \pm y)})
    \nonumber\\
  {P^{m}_{x/y}}&=&<c^{\dagger}_{i m \sigma} c_{j m \sigma}>,
        ~~(j=i \pm \hat{x}/\hat{y})
    \nonumber\\
  {P^{m}_{3}}&=&<c^{\dagger}_{i m \sigma} c_{j m \sigma}>,
        ~~(j=i \pm \hat{(x \pm y)})
\end{eqnarray}
the interaction term in Eq.(4) could be decoupled within the
framework of the mean-field approximation\cite{Seo}. Slightly different from
Seo {$\it et al$.} mean-field ansatz, we consider the contributions of
the kinetic order parameters $P_{x/y}$ and $P_{3}$, since the
itinerant character of the Fe 3d electrons is considerable.
So one could obtain the mean-field Hamiltonian and the ground state
energy.
Minimizing to the ground state energy gives rise to the
self-consistent equations for the order parameters $\Delta$ and $P$,
\begin{eqnarray}
   \Delta^{1 m}_{x/y} &=& \frac{J}{N} \sum_{k}
           cos(k_{x/y})<c^{\dag}_{-k m \uparrow}c^{\dag}_{k m \downarrow}>
    \nonumber\\
   \Delta^{2 m}_{x \pm y} &=& \frac{J^{\prime}}{N} \sum_{k}
           cos(k_{x}\pm k_{y})<c^{\dag}_{-k m \uparrow}c^{\dag}_{k m \downarrow}>
     \nonumber\\
   P^{m}_{x/y}&=& \frac{1}{2N}\sum_{k \sigma} cos(k_{x/y})<n_{k m\sigma}>
     \nonumber\\
   {P^{m}_{3}}&=& \frac{1}{2N}\sum_{k \sigma} cos(k_{x})cos(k_{y})<n_{k m\sigma}>
\end{eqnarray}
The inter-orbital pairing parameter $<c^{\dagger}_{i 1 \uparrow}
c^{\dagger}_{j 2 \downarrow}>$ is neglected due to its very small
value \cite{Seo}. This implies that the $p$-wave pairing symmetry is
precluded.
From Eq.(4), one notices that for each orbital, the hopping along
the $x$-direction is not equivalent to that along the $y$-direction.
This leads to, for example, $\Delta^{1 m}_{x}$$\neq$$\Delta^{1
m}_{y}$.
Due to the equivalence between the $d_{xz}$ and the $d_{yz}$ orbits,
the order parameters of the two orbits are symmetric once
interchanging the $x$-direction and the $y$-direction. Therefore,
throughout this paper, we mainly focus the order parameters in the
first orbit. The results on the second orbit can be arrived at if
rotating the $x$-axis to the $y$-axis \cite{Seo}.
%

Within the present scenario, we could obtain not only the
groundstate phase diagram, but also the quasiparticle spectra in the
normal and the SC states. The temperature dependence of the Fermi
surface in normal state and that of the spin-lattice relaxation rate
in the SC state can also be obtained. Among these quantities, the
spin-lattice relaxation rate in the NMR experiment is expressed as
\cite{Matano}:
\begin{eqnarray}
   \frac{T_{1 N}}{T_{1 s}}=\frac{2}{k_{B} T}
   \int{}{} \int{}{}
N_s(E)N_s(E^{\prime})f(E)[1-f(E^{\prime})\delta(E-E^{\prime})dEdE^{\prime}]
\end{eqnarray}
Providing 1/T$_{1N}$ in the normal state satisfies the Korringa law,
the spin lattice relaxation rate $1/T_{1 s}$ becomes \cite{Xiang}:
\( 1/T_{1 s} \propto  (k_{B} T) \cdot T_{1 N}/T_{1 s} \).
\\

\section{Theoretical Results and Discussions}


\noindent{\bf A. Phase diagram}

\begin{figure}[tp]
\vglue -0.6cm \scalebox{1.150}[1.15]
{\epsfig{figure=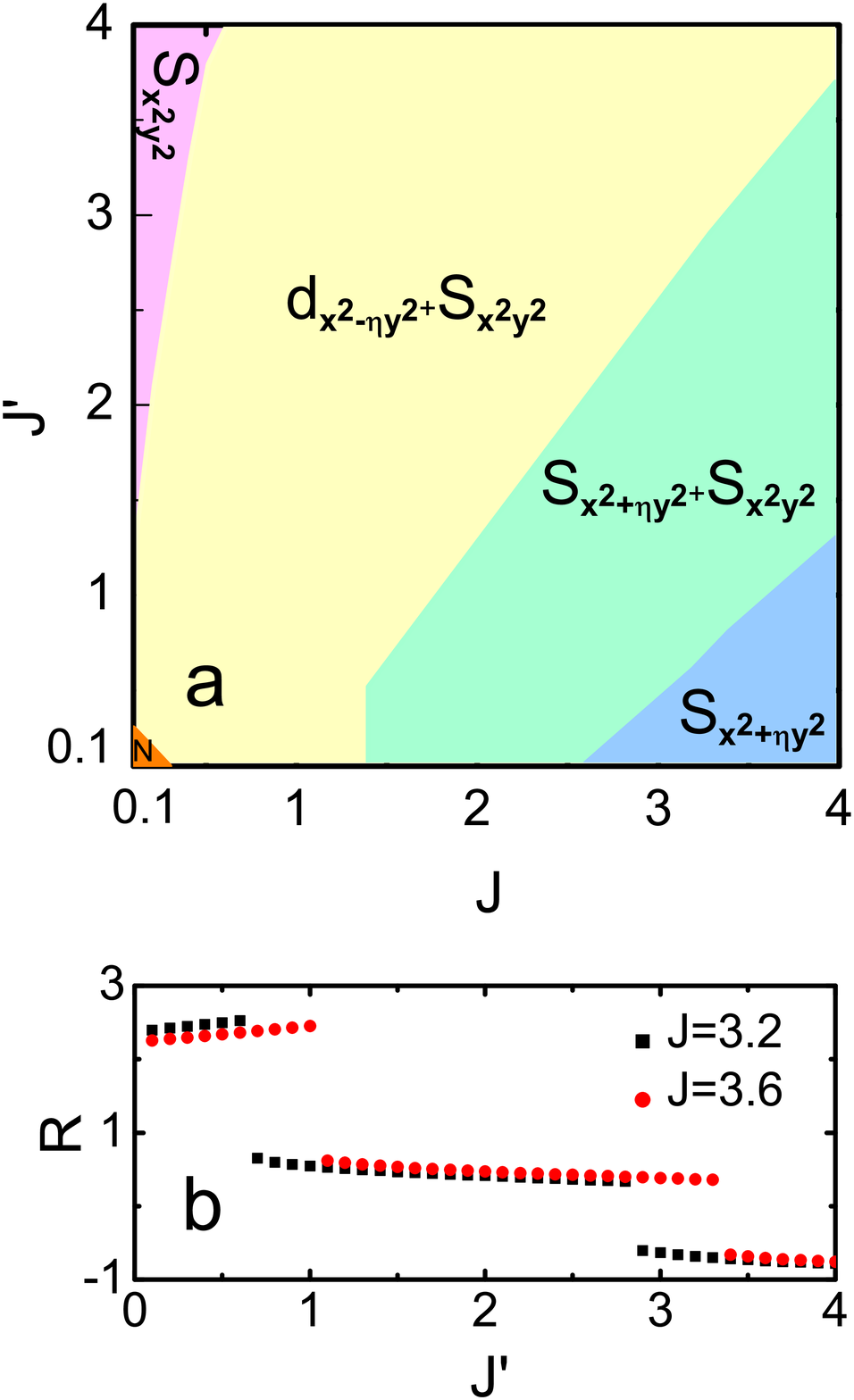,width=7.0cm,angle=0.0}} \caption{(Color
online) The phase diagram of the t-t$^{'}$-J-J$^{'}$ model at
doping concentration x=0.18
from the half-filling. N denotes the normal state, the other four
phases are SC}
\label{fig:fig1}
\end{figure}

Different from Seo {\it et al.}'s phase diagram, we find five stable
phases in our phase diagram, which are shown in Fig.1.
The first one is the normal phase, denoted N in Fig.1. Obviously,
when the superexchange coupling J and J$^{'}$ are too small to
provide the SC pairing glue, the kinetic energy is dominant, and the
electrons stay in the normal state.
Among the SC phases mediated through the spin exchange couplings,
large J and small J$^{'}$ favor the $S_{x_2+\eta y_2}(\propto
cos(kx)+\eta cos(ky))$($\eta >$ 0) SC symmetry, which is the
combination of the $S_{x_2+y_2}$ SC symmetry and $d_{x_2-y_2}$ SC
symmetry. Here $\eta$ is a constant. Different from Seo $et ~al.$
\cite{Seo}, it is not a mixed state of the $S_{x_2+y_2}$ SC state
and the $d_{x_2-y_2}$ one, since the relative phase of the
coefficients of the $S_{x_2+y_2}$ component and the $d_{x_2-y_2}$
one is fixed, hence it is a pure state.
%
%
On the other hand, small J and large J$^{'}$ favors the $S_{x_2y_2}
\propto cos(kx+ky)+cos(kx-ky)$ SC phase, since the NNN hopping
integrals along the $x$-axis and the $y$-axis are identical.
As seen in Fig.1a, in the regions where J and J$^{'}$ compete with
each other, the third SC phase appears as the combination of the
$S_{x_2+\eta y_2}$ and the $S_{x_2y_2}$ symmetries.
%
%
For the case in this region with $J=3$ and $J^{'}=1.5$, the SC order
parameters are $\Delta^{1 1}_{x}$=0.241, $\Delta^{1 1}_{y}$=0.555
and $\Delta^{2 m}_{x \pm y}$=0.060.
%
%
%
And the fourth SC phase appears with the combination of the
$d_{x_2-\eta y_2}(\propto cos(kx)-\eta cos(ky))$($\eta >$ 0) and
$S_{x_2y_2}$ symmetries.
%
%
For the case in this region with J=0.7 and $J^{'}$=0.3, the SC order
parameters are $\Delta^{1 1}_{x}$=-0.039, $\Delta^{1 1}_{y}$=0.029,
and $\Delta^{2 m}_{x \pm y}$=0.012.
%
%
The components of the SC order parameters strongly depend on the
relative magnitude of J$^{'}$ with respect to J in different SC
regions. As we can see the ratio $\Delta^{1 1}_{x}/\Delta^{1 1}_{y}$
=R in Fig.1b, the change of the ratio R is discontinuous with the
increase of $J^{'}$ at fixed $J$, implying that the quantum phase
transitions between these SC phases are the first order.
\\

\noindent{\bf B. ARPES and T-Dependence of SC Gaps}
\begin{figure}[tp]
\vglue -0.6cm \scalebox{1.150}[1.15]
{\epsfig{figure=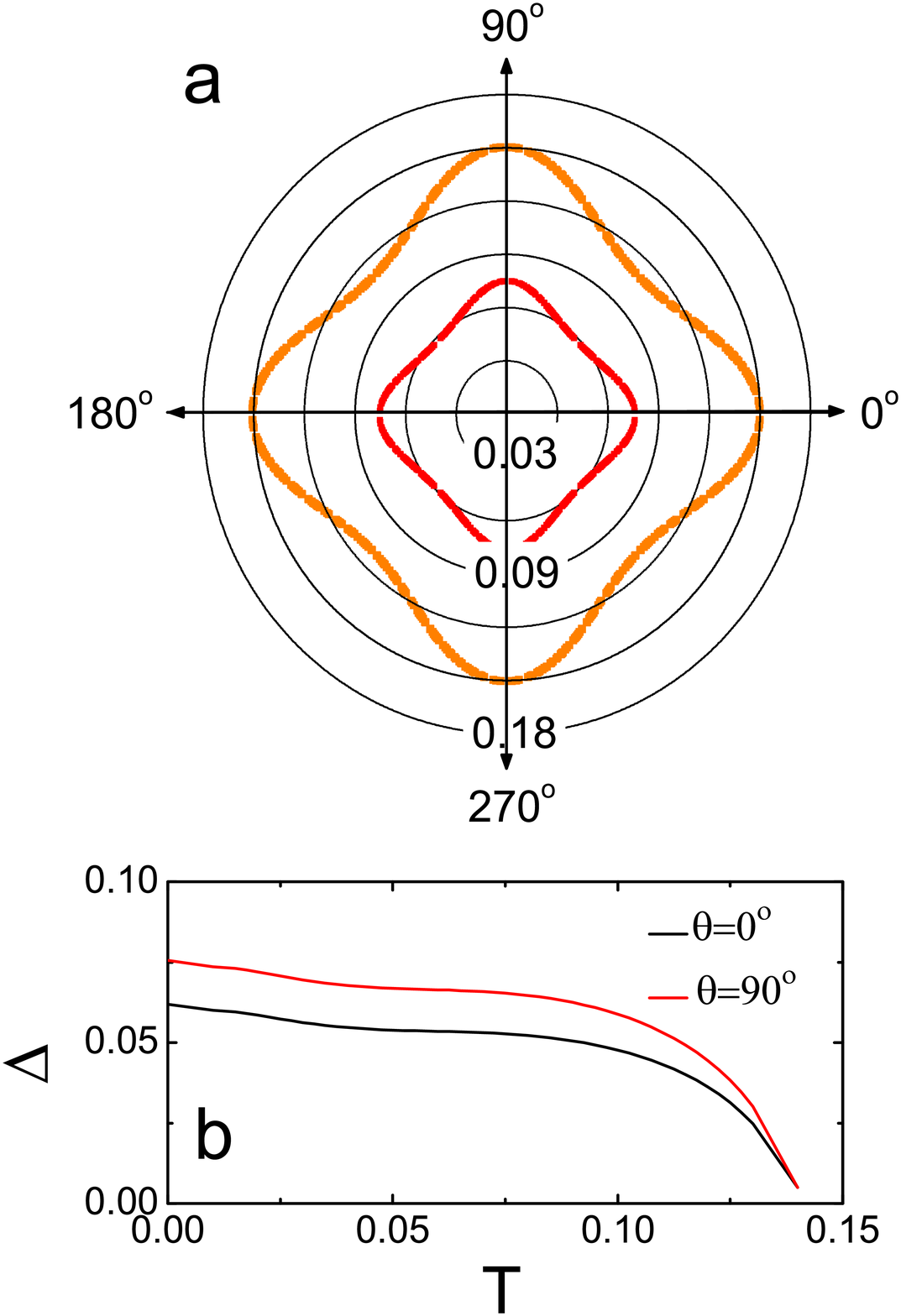,width=7.0cm,angle=0.0}} \caption{(Color
online)The angle dependence of the SC gaps near the small hole FS
(red line) and the large hole FS (orange line) around the $\Gamma$
point in the polar coordinates. Theoretical parameters: $J=0.7$,
$J^{'}=0.3$, and doping concentration x=0.18.}
\label{fig:fig2}
\end{figure}

The angle-resolved photoemission spectra (ARPES) experiment provides
direct information about the quasiparticle spectra in normal state
and the symmetry of the SC gaps in the SC phase. We present the SC
gap character of the t-t$^{'}$-J-J$^{'}$ models for the parameters
in LaFeAsO in Fig. 2. Notice that the benchmark of the Fermi surface
adopted here to plot the SC gap is at $T=0.2$.
Our results show that the SC gap structure exhibits fourfold
symmetry, as seen in Fig.2a, in accordance with the Fermi surface
topology of the present model. The amplitudes of the SC gaps on the
different Fermi surfaces are different, suggesting the nature of a
two-gap SC.
The SC gap structures are anisotropic and nodeless, as shown in
Fig.2a. Though the SC gap structure likes an anisotropic $s-$wave
pairing symmetry, it is in fact the symmetry of the $d_{x^{2}-\eta
y^{2}}$+$S_{x^{2}y^{2}}$-wave. Our result reaches an agreement with
recent ARPES experiments by Kondo $et al.$ \cite{Kondo}.
We also notice that the relative variation of the anisotropic gaps
is less than 25 $\%$, so our results do not conflict with Ding $et
~al.'s $ \cite{Ding} and Zhou $et ~al.$'s \cite{Zhou} reports.
%
%

Fig.2b shows the temperature dependence of the SC energy gap
$\Delta$ along the $\theta=0$ and $\theta=90\deg$. With the
increasing of the temperature, the SC order parameter in the small
hole-like FS sheets decrease monotonously, as observed in the ARPES
experiments \cite{Ding,Zhou}.
Near $T=0.04$, both the SC gaps exhibit a small dip, which is
attributed to the interactions among the three SC energy scales.
Obviously, the nodeless and the anisotropy of the SC gaps originate
from the unique gap structure, $\Delta^{m}= \Delta^{2 m}_{x \pm y}
(\cos k_x \cos k_y) + (\Delta^{1 m}_{x}\cos k_x- \Delta^{1
m}_{y}\cos k_y)$. And such a symmetry is also in agreement with the
analysis on the SC pairing symmetry \cite{WangQH,LinHQ}.
%
%
However, one finds that the magnitude of the gap in the large FS is
bigger than that in small FS in the $\Gamma$ point, opposite to the
ARPES result. This may be due to the tight-binding parameters of the
two-orbital t-t$^{'}$-J-J$^{'}$ model, which only describes the
topology structure of the Fermi surfaces of the FeAs
superconductors, and does not contain all the details of the Fermi
surfaces and the band structures in ironpnictide compounds. Hence,
we expect that the refined tight-binding parameters will improve our
results in the further study.
\\

\noindent{\bf C. T-dependence of Fermi Surface}
\begin{figure}[tp]
\vglue -0.6cm \scalebox{1.150}[1.15]
{\epsfig{figure=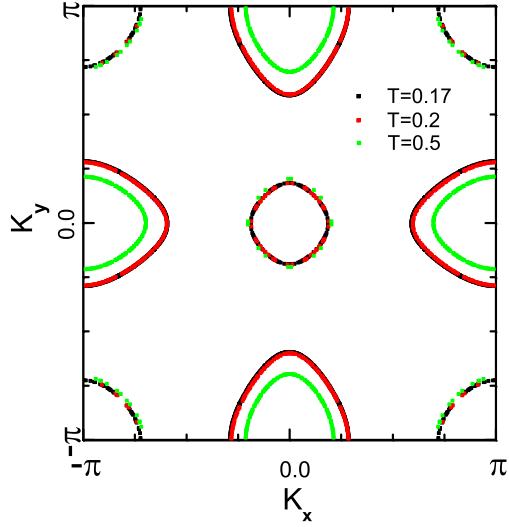,width=7.0cm,angle=0.0}} \caption{(Color
online) (a) The Fermi surface for the different temperature, T=0.17,
0.2, 0.5. (b) The theoretical parameters are the same as the Fig.2.
}
\label{fig:fig3}
\end{figure}

The temperature evolution of the FS in the normal state is shown in
Fig. 3, when T is higher than the critical temperature of the SC
phase transition.
The two hole-like and electron-like FS sheets can be clearly
identified around the $\Gamma$ and the $M$ points at x= 0.18, as
observed in the ARPES experiments and the first-principle electronic
structures calculations.
Interestingly, the hole-like FS sheets expand a little with the
increasing of the temperature. In contrast, the electron-like FS
sheets shrink very acutely. This may demonstrate that the
electron-like FS sheets are more important for the occurrence of the
SC state, in consistent with Dai $et al.$ prediction. \cite{Xi}.
%
%
%
\\

\noindent{\bf D. Spin-Lattice Relaxation Rate in NMR}
\begin{figure}[tp]
\vglue -0.6cm \scalebox{1.150}[1.15]
{\epsfig{figure=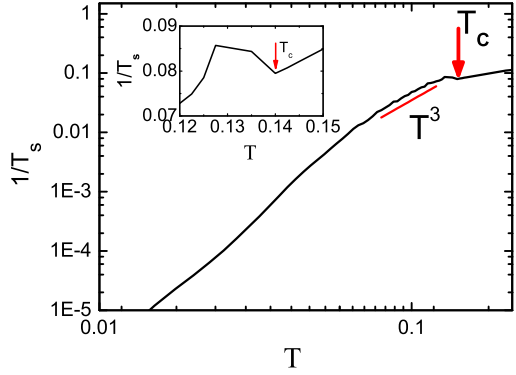,width=7.0cm,angle=0.0}} \caption{(Color
online) Temperature dependence of the spin lattice relaxation rate
in the t-t$^{'}$-J-J$^{'}$ model. The red arrow indicates the SC
critical temperature T$_{c}$. The red line is the T$^{3}$ law for
comparison. Inset shows the detail near T$_{c}$. The theoretical
parameters are the same as in Fig.2.}
\label{fig:fig4}
\end{figure}

Although many experimental measurements, such as the Andreev
reflection \cite{PCAR1}, the exponential temperature dependence of
the penetration depths \cite{pd1} and the ARPES \cite{Zhou,Ding}
observe the nodeless gap function in the SC phase of
ReFeAsO$_{1-x}$F$_{x}$ and Ba$_{1-x}$K$_{x}$Fe$_{2}$As$_{2}$
compounds,
the line nodes in the SC gap was also suggested by the NMR
experiment\cite{Matano}. The two characters in the NMR experiment
supported the line nodes: lack of the coherence peak and the $T^{3}$
behavior in the nuclear spin-lattice relaxation rate, $1/T_{1}$.
Using the gap function obtained in this paper, we calculate the spin
lattice relaxation rate 1/T$_{s}$, and the numerical result
%
%
is shown in Fig. 4. We also plot the $T^{3}$ law (the red line) for
a comparison. It is found that over a wide temperature range, the
spin lattice relaxation rate in the present model can be fitted by
the $T^{3}$ law, in agreement with the observation of the NMR
experiments \cite{Matano,Grafe}.

A small coherence peak appears around the critical transition
temperature, as clearly seen in the inset of Fig.4. Experimentally,
such a small coherence peak may be easily suppressed by the impurity
effect or the antiferromagnetic spin fluctuations, similar to the
situations in cuprates. This leads to the missing of the
Hebel-Slichter coherence peak in the NMR experiment in ironpnictide
SC.
With the decreasing of the temperature, one find a drop in the spin
lattice relaxation rate, $1/T_{1s}$, consistent with the observation
of the NMR experiments \cite{Matano}.
Such a behavior deviating from the $T^{3}$ law may contribute from
the multi-gap character of this system, and such a drop reflects the
different SC gaps in different orbits. Surely, more meticulous
studies are needed in near future.
We also notice that Parker $et al.$ found the extended $s_{\pm}$ SC
gaps also can give the same NMR relaxation rate in SC pnictides
\cite{Parker}. The inter-band contribution to the spin-lattice
relaxation deviating from the $T^{3}$ law was suggested by Parish $et
~al. $ \cite{Parish}.

  Therefore, we find that the numerous unusual properties in the normal
state and the SC phase of newly discovered FeAs superconductors
could be interpreted in the t-t$^{'}$-J-J$^{'}$ model, showing that
this model is a good approximate model to describe the iron-based
superconductors. Within this scenario, the superimposition and
mixing of the pairing electrons in the two orbits contributes to the
anisotropic nodeless SC pairing symmetry.
Such a pairing symmetry assembles the characters of usual $d$-wave
and $s$-wave, hence shares the properties of the usual $d$-wave SC,
like cuprates, and the $s$-wave SC, such as MgB$_{2}$ \cite{Zhou}.
Nevertheless, to compare the theoretical results with the
experimental observation quantitatively, more subtle band structures
in the t-t$^{'}$-J-J$^{'}$ model are needed, though the present
two-orbital model gives the qualitatively correct properties. Also
the present constrained mean-field approximation should be improved
in further study.

\section{Summary}

In summary, starting with the t-t$^{'}$-J-J$^{'}$ model, we obtain the
mean-field parametric phase diagram at the doping concentration
x=0.18, and find the normal state and a new SC phase in the phase
diagram, different from the literature.
With the decrease of the temperature, a nodeless and anisotropic
$d_{x^{2}-\eta y^{2}}$+$S_{x^{2}y^{2}}$-wave gap structure emerges.
But due to the multi-gap character, the anisotropic gaps open on the
hole Fermi surfaces and the electron Fermi surfaces are different.
The T$^{3}$ law of the spin lattice relaxation rate, 1/T$_{1}$, is
also interpreted in the present t-t$^{'}$-J-J$^{'}$ model.
%

%
\acknowledgments

This work was supported by the NSFC of China no.90303013, the BaiRen
Project and the Knowledge Innovation Program of Chinese Academy of
Sciences. Part of the calculations were performed in Center for
Computational Science of CASHIPS and the Shanghai Supercomputer
Center.

\bibliography{apssamp}

\end{document}